\pdfoutput=1
\documentclass[11pt]{article}
\usepackage[margin=1in]{geometry}

\usepackage{amsmath,amsfonts}
\usepackage{graphicx}
\usepackage{tikz}
\usetikzlibrary{arrows.meta,calc}
\usepackage{array}
\usepackage{circledsteps}
\usepackage{booktabs}
\usepackage{caption}
\usepackage{subcaption}
\usepackage{url}
\usepackage{verbatim}
\usepackage{cite}
\usepackage{algorithm}
\usepackage{algpseudocode}
\usepackage{dsfont}
\usepackage{xcolor}
\usepackage{authblk}
\usepackage[hidelinks]{hyperref}

\begin{document}

\title{EMT Modeling and Coordinated Mode-Transition Ride-Through Control of Modern Data Centers}

\author[1,*]{Junyuan Zheng}
\author[2,*]{Salish Maharjan}
\author[2]{Min Lwin}
\author[2]{Soummya Roy}
\author[1]{Zhaoyu Wang}
\author[1]{Wenlong Shi}
\author[3]{Feng Qiu}
\author[4]{Rajarshi Roychowdhury}
\affil[1]{Iowa State University, IA, USA \protect\\ \{\texttt{zhengjy, wshi5, wzy}\}\texttt{@iastate.edu}}
\affil[2]{INS Engineering, TX, USA \protect\\ \{\texttt{smaharjan, mlwin, sroy}\}\texttt{@ins-engineering.com}}
\affil[3]{Argonne National Laboratory, IL, USA \protect\\ \texttt{fqiu@anl.gov}}
\affil[4]{Xcel Energy, CO, USA \protect\\ \texttt{raj.roychowdhury@xcelenergy.com}}
\affil[*]{Junyuan Zheng and Salish Maharjan contributed equally to this work and share first authorship.}
\date{}

\setlength{\abovedisplayskip}{3.5pt}
\setlength{\belowdisplayskip}{3.5pt}

\maketitle

\begin{center}
\fbox{\parbox{0.92\linewidth}{\centering\small
This work has been submitted to the IEEE for possible publication. Copyright may be transferred without notice, after which this version may no longer be accessible.}}
\end{center}

\begin{abstract}
Large data centers are emerging as power-electronic loads whose response to grid voltage disturbances can affect bulk-system stability. This paper presents the composition and control characteristics of a proprietary electromagnetic transient (EMT) model of a modern data center comprising an uninterruptible power supply (UPS) with battery energy storage, induction-motor cooling loads, a computer room air handler (CRAH) driven by a variable-frequency drive (VFD), and an aggregated server load. A voltage-based state machine coordinates three aspects of the facility response: the UPS operating mode (eco, double conversion, or battery), the supply path of the CRAH, and a ramped reconnection of the grid-side UPS converter after the disturbance clears. A 100~MW facility is implemented in PSCAD/EMTDC and benchmarked against the PSS\textregistered E PERC1 aggregate data center model under a symmetrical three-phase fault and low- and high-voltage ride-through (LVRT/HVRT) events. The two models agree closely in their voltage and active-power trajectories, while the EMT model also resolves the converter and motor-reacceleration transients that the phasor-domain model cannot.
\end{abstract}


\section{Introduction}

Artificial intelligence (AI) workloads are turning data centers into one of the fastest-growing classes of electric load. Global data center electricity consumption is projected to more than double, from about 415~TWh in 2024 to about 945~TWh by 2030, or nearly 3\% of global electricity demand~\cite{iea2025}. In the United States, data centers used 4.4\% of total electricity in 2023, and that share could reach 6.7--12\% by 2028~\cite{lbnl2024}. Single campuses now request transmission-level interconnections of hundreds of megawatts, so the dynamic behavior of one facility is no longer negligible in bulk-system studies.

Recent events show why this matters. In July 2024, a 230~kV line fault in the Eastern Interconnection caused about 1,500~MW of data center load to transfer to backup power at the same time. Operators had not anticipated the load loss, and system voltages rose above normal operating levels~\cite{nerc2025incident}. The NERC Large Loads Task Force has since identified voltage sensitivity, rapid demand swings, and coordinated load loss of this kind as stability risks that current planning practice does not cover~\cite{nerc2025lltf}. What the grid sees during such an event is not a fixed load characteristic. It is set by the facility's own controls: millisecond-scale voltage thresholds and timers that move the UPS between operating modes, under-voltage protection on the cooling motors, and the rate at which the UPS rebuilds its grid draw after the fault clears.

Phasor-domain models such as the PSS\textregistered E PERC1 data center model and the WECC composite load model (CMPLDWG)~\cite{psse,kosterev2008} are efficient and widely used in planning. However, they represent the facility only at its aggregate terminals with half-cycle time steps. They cannot resolve the dynamics that decide whether a facility rides through or drops off: sub-cycle voltage detection, DC-link and converter control transients, the handover from rectifier to battery converter, and induction-motor stalling and reacceleration. These effects become more important on weak grids with high shares of inverter-based resources, where system operators increasingly rely on electromagnetic transient (EMT) studies. EMT models of data centers have only recently become available. The Pacific Northwest National Laboratory (PNNL) data center model library offers generic EMT components and aggregated facility designs, but these have not yet been validated against measured data~\cite{pnnl2026}, and ERCOT has released a PSCAD model of an AI data center load~\cite{ercot2026}. To the best of our knowledge, no published EMT model represents the facility-level logic that coordinates UPS modes, cooling-system supply paths, and post-fault reconnection, and benchmarks the result against the phasor models used in interconnection studies.

This paper addresses that gap. Its main contributions are as follows:
\begin{enumerate}
\item A scalable, parameterized EMT model of a grid-connected data center built in PSCAD/\allowbreak EMTDC. It includes a UPS with rectifier, inverter, bidirectional DC--DC converter, and battery energy storage; induction-motor chiller and cooling-tower loads with under-voltage trip and reclose; a VFD-driven computer room air handler (CRAH) with a dual supply path; and a data-driven aggregated server load.
\item A voltage-based four-state machine that coordinates the UPS operating mode (eco, double conversion, or battery), the CRAH supply path, and a ramped reconnection of the UPS rectifier. Hysteresis bands and a ramp-abort rule prevent mode chattering and repeated full-power reconnection during successive voltage dips.
\item A benchmark of a 100~MW facility against the PSS\textregistered E PERC1 and CMPLDWG models under a three-phase fault and LVRT and HVRT events. The models agree closely at the aggregate terminals, and the EMT model additionally resolves converter and motor-reacceleration transients. The results also show that a successful ride-through appears to the grid as a sudden loss of most of the facility demand, followed by a controlled ramp back.
\end{enumerate}

The rest of the paper is organized as follows. Section~\ref{II} describes the data center composition and its operating modes. Section~III presents the simulation results, and Section~IV concludes the paper.

\begin{figure}[t]
  \centering
  \captionsetup[subfloat]{justification=centering,singlelinecheck=false}

  \subfloat[]{%
    \includegraphics[width=0.85\linewidth]
    {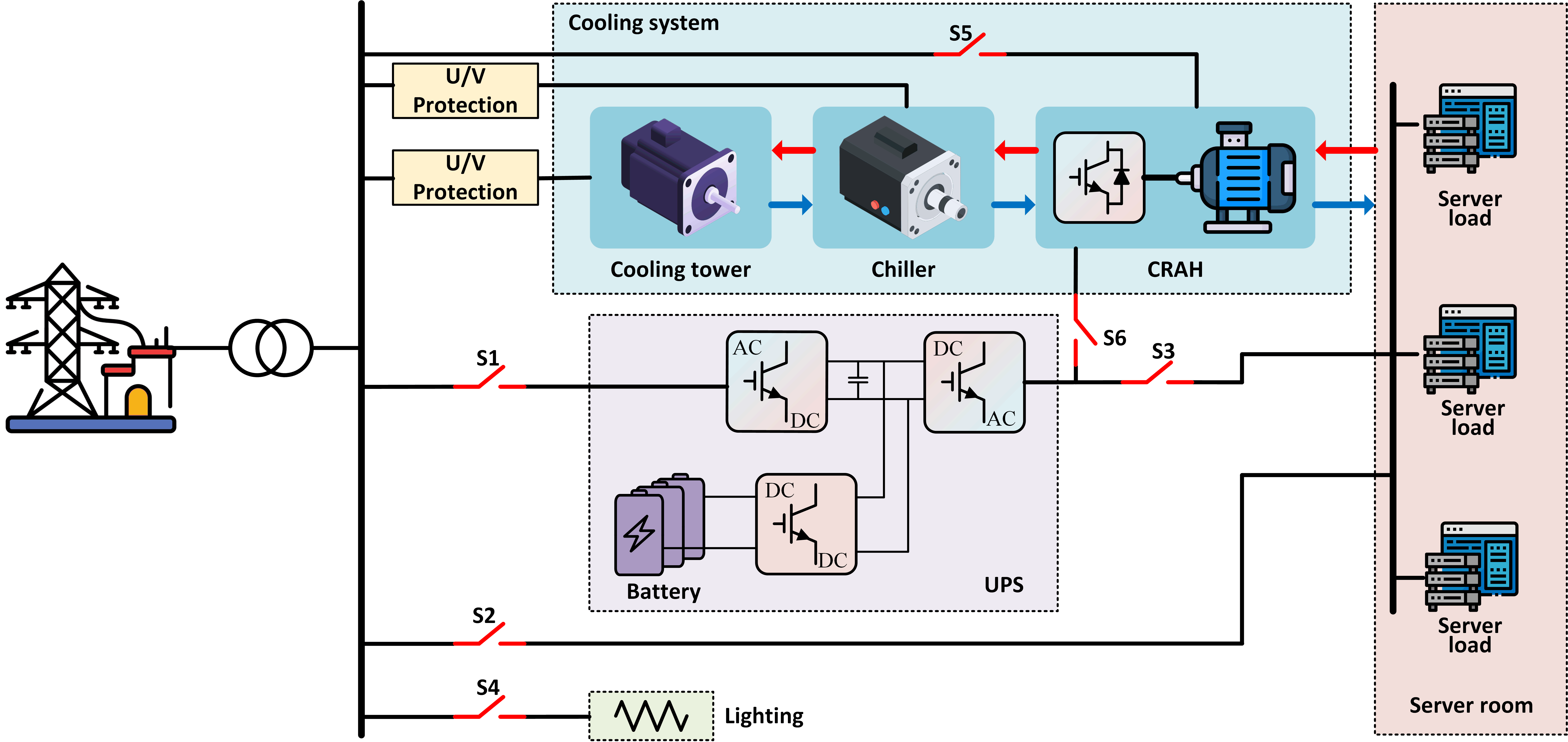}%
    \label{dc_structure}%
  }

  \vspace{2mm}

  \subfloat[]{%
    \includegraphics[width=0.5\linewidth]{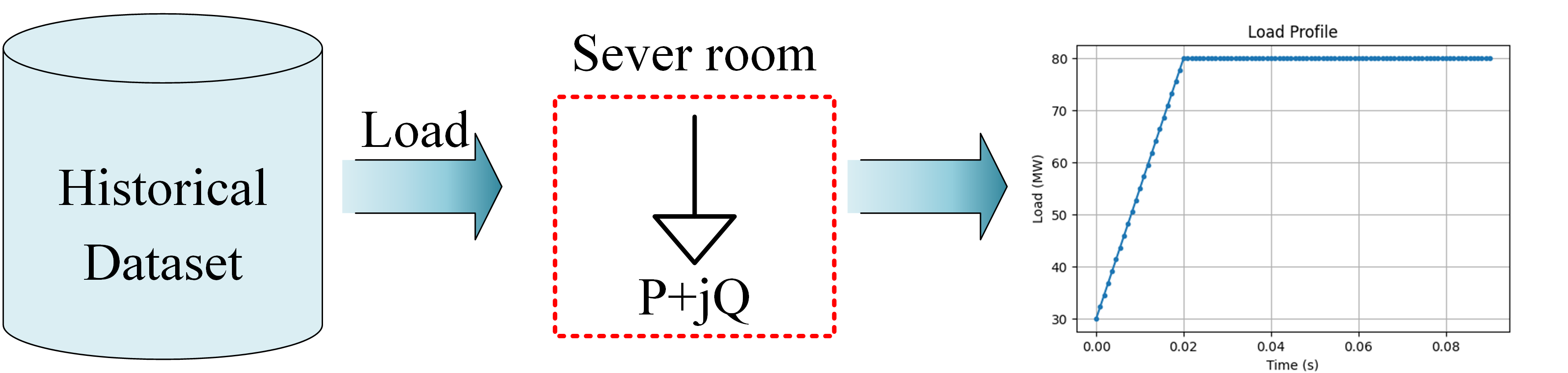}%
    \label{server_load}%
  }

  \caption{Data center configuration and server load modeling:
  (a) AC power distribution system;
  (b) data-driven server load modeling framework.}
  \label{fig:dc_framework}
\end{figure}

\section{Modern Data Center Composition And Operation}\label{II}
Fig.~\ref{dc_structure} shows a typical subsystem configuration of a modern data center. It comprises (a) a cooling system, (b) an uninterruptible power supply (UPS), (c) server/IT loads, and (d) lighting and auxiliary loads. These subsystems work together to ride through grid disturbances and meet grid interconnection requirements while keeping the server load continuously supplied. Internal control details are omitted because of proprietary restrictions.
\subsection{Composition}
\subsubsection{Cooling tower and chiller}
The chiller compressor produces chilled water that removes heat from the server room through the chilled-water loop, while the cooling tower uses large fans to reject heat from the condenser-water loop to the ambient environment. The compressor and the fans are typically driven by three-phase induction motors. These motors are connected directly to the grid and are tripped by under-voltage protection during grid events. Both the chiller and the cooling tower are therefore modeled as aggregated three-phase induction motors based on the WECC composite load model (CMPLDWG)~\cite{kosterev2008}. Since CMPLDWG is available only as a built-in PSS\textregistered E model, we implemented an equivalent model in PSCAD, including its under-voltage trip and reclose scheme.

\subsubsection{CRAH}
The CRAH circulates air through the server room to remove heat and maintain the required thermal environment. It is modeled as a two-stage VFD made up of an AC--DC rectifier, a DC--AC inverter, and a three-phase induction motor that drives the fan. The rectifier regulates the DC-link voltage and the input reactive power, while the inverter uses V/f control to produce variable-frequency AC power and set the motor speed. Because the CRAH is an essential load, it can be supplied either from the AC grid through switch $S_5$ or from the UPS through switch $S_6$. Under normal conditions, $S_5$ is closed and $S_6$ is open, so the CRAH is fed directly from the grid. During a severe grid disturbance, $S_5$ opens and $S_6$ closes, transferring the CRAH to the UPS.

\subsubsection{UPS}
The UPS keeps critical loads supplied during grid disturbances such as voltage sags and short interruptions. It consists of an AC--DC rectifier, a DC--AC inverter, a battery energy storage system (BESS), and a bidirectional DC--DC converter. The rectifier converts grid-side AC power into DC and regulates the DC-link voltage. The inverter converts the DC power back into AC to supply the server load, and the bidirectional DC--DC converter controls battery charging and discharging. At a high level, the DC--DC converter is a buck--boost converter that regulates power in buck mode and the DC-link voltage in boost mode. The battery provides the backup energy that keeps the server load and the CRAH supplied during grid disturbances.

\subsubsection{Server Loads}
As illustrated in Fig.~\ref{server_load}, the server room is modeled with a data-driven approach as a single aggregated load. Measured operational data are used to build a time-series load profile that captures how server demand varies over time.

\subsubsection{Lighting loads}
Lighting loads serve auxiliary functions such as room illumination, maintenance access, and safety. They are modeled as static constant-power loads.

\subsection{Operating Modes and Mode-Transition Logic}
\label{sec:modes}

\begin{figure}[!t]
\centering
\begin{tikzpicture}[
  font=\footnotesize,
  st/.style={draw, rounded corners=3pt, align=center, inner sep=3pt,
             text width=2.55cm, minimum height=1.25cm, fill=gray!8},
  tr/.style={-{Stealth[length=2mm]}, thick},
  lb/.style={font=\footnotesize, fill=white, inner sep=1pt}
]
\node[st] (M0) at (0,0)      {\textbf{M0: Normal eco}\\ UPS bypass; IT \& CRAH\\ on grid};
\node[st] (M1) at (4.45,0)    {\textbf{M1: Double conv.}\\ IT via UPS;\\ CRAH on grid};
\node[st] (M2) at (0,-3.0)   {\textbf{M2: Battery}\\ IT \& CRAH on BESS;\\ $P_g^{\mathrm{UPS}}=P_{\mathrm{cease}}$};
\node[st] (M3) at (4.45,-3.0) {\textbf{M3: Recovery}\\ $P_g^{\mathrm{UPS}}$ ramps\\ per (\ref{eq:ramp1})};

\draw[tr] (M0.20)  -- node[lb, above]{$C_{\mathrm{DC}}$} (M1.160);
\draw[tr] (M1.200) -- node[lb, below]{$C_{\mathrm{N}}$}  (M0.340);
\draw[tr] (M0) -- node[lb, left]{$C_{\mathrm{B}}$} (M2);
\draw[tr] (M1.south west) -- node[lb, pos=0.45]{$C_{\mathrm{B}}$} (M2.north east);
\draw[tr] (M2.340) -- node[lb, below]{$C_{\mathrm{R}}$ + $T_{\mathrm{start}}$} (M3.200);
\draw[tr] (M3.160) -- node[lb, below, pos=0.6]{$C_{\mathrm{B}}$} (M2.20);
\draw[tr] (M3) -- node[lb, right]{$C_{\mathrm{D}}$} (M1);

\draw[tr] ($(M0.west)+(-0.55,0)$) -- (M0.west);
\end{tikzpicture}
\caption{Mode-transition state machine of the data center. Transition
conditions $C_{\mathrm{B}}$, $C_{\mathrm{DC}}$, $C_{\mathrm{R}}$,
$C_{\mathrm{N}}$, and $C_{\mathrm{D}}$ are defined in
(\ref{eq:CB})--(\ref{eq:CD}). $C_{\mathrm{B}}$ is evaluated in every state
and has priority.}
\label{fig:fsm}
\end{figure}
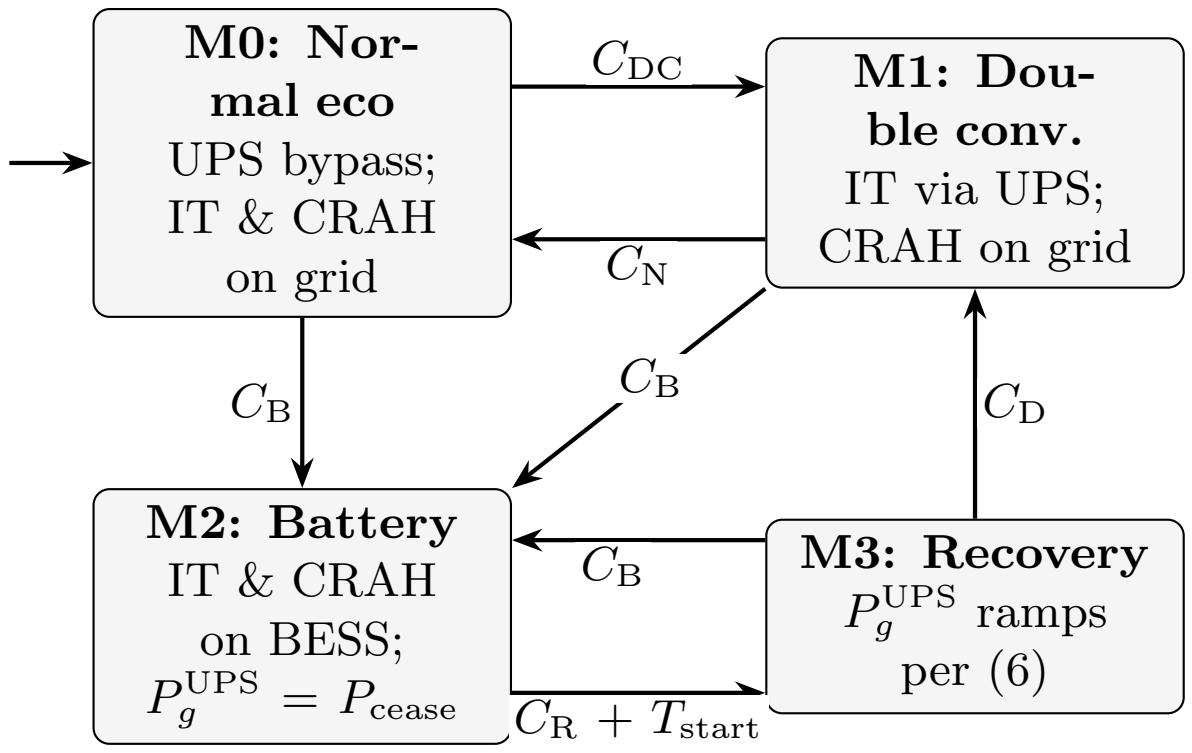

The facility control aims to keep the server load continuously supplied during grid disturbances while shaping the power drawn from the grid during and after the event. The facility has four operating modes: eco, double conversion, battery, and recovery. Their coordination is implemented as the four-state machine shown in Fig.~\ref{fig:fsm}, driven by the per-unit grid voltage $V_t$. The thresholds and timers, listed in Table~\ref{tab:params}, are chosen to be consistent with ISO interconnection requirements and with the voltage-tolerance envelope of server power supplies~\cite{itic}. The switch states in each mode are given in Table~\ref{tab:modes}.

The transition conditions are defined as
\begin{align}
C_{\mathrm{B}}  &: \big(V_t < V_{\mathrm{LV}} \text{ for } T_{\mathrm{LV}}\big)
   \lor \big(V_t > V_{\mathrm{HV}}\text{ for } T_{\mathrm{HV}}\big)
, \label{eq:CB}\\
C_{\mathrm{DC}} &: V_{\mathrm{LV}} \le V_t <  \underline{V_{\mathrm{N}}} \text{ for } T_{\mathrm{DC}}, \label{eq:CDC}\\
C_{\mathrm{R}}  &: V_{\mathrm{LVrecon}} \le V_t \le V_{\mathrm{HVrecon}} \text{ for } T_{\mathrm{recon}}, \label{eq:CR}\\
C_{\mathrm{N}}  &: \underline{V_{\mathrm{N}}} \le V_t \le \overline{V_{\mathrm{N}}} \text{ for } T_{\mathrm{recon}}+T_{\mathrm{ramp}}, \label{eq:CN}\\
C_{\mathrm{D}}  &: P_g^{\mathrm{UPS}}(t) = P_{\mathrm{nom}} \;\; (t \ge T_1+T_{\mathrm{ramp}}), \label{eq:CD}
\end{align}
where $P_g^{\mathrm{UPS}}$ is the active power drawn from the grid by the UPS rectifier, and $\underline{V_{\mathrm{N}}}$ and $\overline{V_{\mathrm{N}}}$ are the lower and upper voltage limits within which the facility operates in normal eco mode. The remaining variables are defined in Table~\ref{tab_mod_transition}. Each timer resets whenever its voltage condition is violated. $C_{\mathrm{B}}$ has priority over all other conditions and is evaluated in every state. Separate thresholds for entering and leaving each mode provide hysteresis, which prevents chattering during voltage recovery.

\subsubsection{M0 -- Normal eco mode}
For $\underline{V_{\mathrm{N}}} \le V_t \le \overline{V_{\mathrm{N}}}$, the server load and the CRAH are supplied directly from the grid, and the UPS operates in bypass through $S_2$. The rectifier remains energized through $S_1$ to keep the DC link charged and the battery at its target state of charge. Short sags that satisfy neither $C_{\mathrm{DC}}$ nor $C_{\mathrm{B}}$ are seen directly by the server power supplies, so the timers are set inside the voltage-tolerance envelope.

\subsubsection{M1 -- Double-conversion mode}
If $C_{\mathrm{DC}}$ holds, $S_3$ closes and $S_2$ opens, and the server load is supplied through the rectifier--inverter path. The rectifier regulates the DC-link voltage, and the inverter regulates the server-bus voltage. The CRAH and the directly connected chiller and cooling-tower motors remain on the grid. When $C_{\mathrm{N}}$ holds, the facility returns to M0: $S_3$ opens first, and $S_2$ then closes with as little delay as possible.

\subsubsection{M2 -- Battery mode}
If $C_{\mathrm{B}}$ holds in any state, $S_1$ opens and isolates the UPS from the grid. The bidirectional DC--DC converter then regulates the DC link from the BESS, and the UPS inverter operates in grid-forming mode to supply the server load and, through $S_6$, the CRAH.

\subsubsection{M3 -- Ramped recovery}
\begin{figure}[t]
  \centering
  \includegraphics[width=0.5\linewidth]{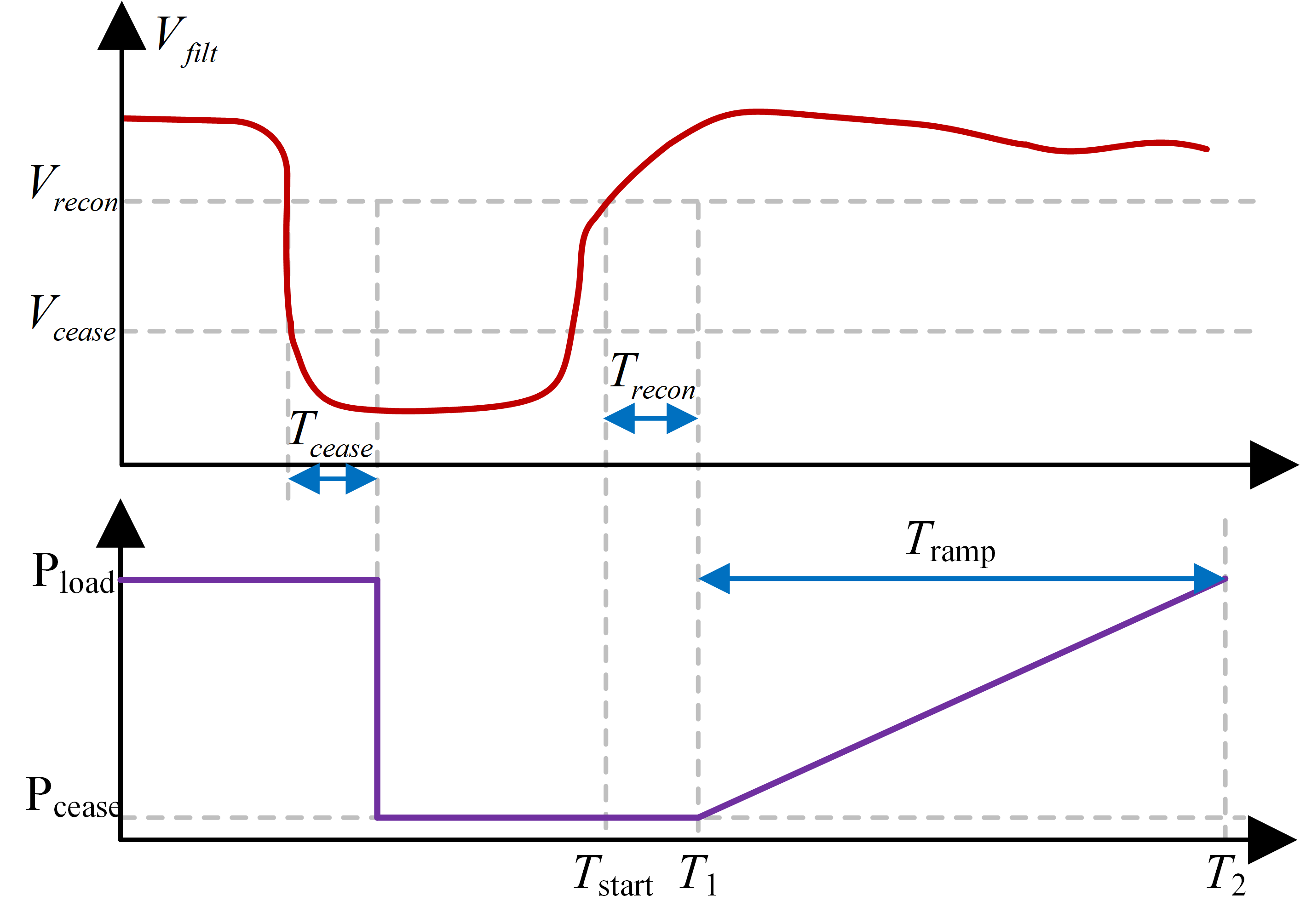}
  \caption{Cease and reconnect logic of the data center.}
  \label{fig_cease_logic}
\end{figure}

When $C_{\mathrm{R}}$ has held for $T_{\mathrm{recon}}$, $S_1$ recloses and the rectifier switches to power-control mode. Let $T_\mathrm{start}$ denote the instant at which $V_t$ re-enters the reconnection band. The ramp then starts at $T_1 = T_\mathrm{start} + T_{\mathrm{recon}}$, as shown in Fig.~\ref{fig_cease_logic}, and the rectifier power reference follows
\begin{align}
P_g^{\mathrm{UPS}}(t) &= P_{\mathrm{cease}}
  + \frac{t - T_1}{T_{\mathrm{ramp}}}\big(P_{\mathrm{nom}} - P_{\mathrm{cease}}\big), \label{eq:ramp1}
\end{align}
for $T_1 \le t < T_1 +T_\mathrm{ramp}$, where $P_{\mathrm{nom}}$ is the pre-disturbance UPS input power and $P_{\mathrm{cease}}$ is the grid-side UPS power at the start of the ramp. $P_{\mathrm{cease}}$ is generally zero because the UPS is disconnected from the grid before recovery begins. Throughout M3, the DC--DC converter regulates the DC link. The BESS therefore supplies the difference between the IT demand and $P_g^{\mathrm{UPS}}(t)$, and the server load remains unaffected by the ramp. When the ramp completes ($C_{\mathrm{D}}$), the rectifier resumes DC-link control and the CRAH is transferred back to the grid ($S_6$ opens, $S_5$ closes). The facility then enters M1 and returns to M0 only after $C_{\mathrm{N}}$ is satisfied. If $C_{\mathrm{B}}$ occurs during M3, the ramp is aborted and the facility returns to M2. This avoids repeated full-power reconnection during successive voltage dips, for example during line auto-reclosing.

\begin{table}[!t]
\centering
\caption{Switch states in each operating mode}
\label{tab:modes}
\footnotesize\setlength{\tabcolsep}{4pt}
\begin{tabular}{llcccccc}
\hline
Mode & Description & $S_1$ & $S_2$ & $S_3$ & $S_4$ & $S_5$ & $S_6$ \\
\hline
M0 & Normal eco           & ON  & ON  & OFF & ON & ON  & OFF \\
M1 & Double conv.         & ON  & OFF & ON  & ON & ON  & OFF \\
M2 & Battery              & OFF & OFF & ON  & ON & OFF & ON  \\
M3 & Recovery             & ON  & OFF & ON  & ON & OFF & ON  \\
\hline
\end{tabular}
\end{table}

\begin{table}[!t]
\centering
\caption{Mode-transition thresholds and timers}\label{tab_mod_transition}
\label{tab:params}
\footnotesize\setlength{\tabcolsep}{4pt}
\begin{tabular}{lll}
\hline
Quantity & Threshold & Timer \\
\hline
Normal voltage limits ($\underline{V_{\mathrm{N}}}$, $\overline{V_{\mathrm{N}}}$) & (0.9 pu, 1.15 pu) & $\infty$\\
Low voltage cease ($V_{\mathrm{LV}}$, $T_{\mathrm{LV}}$)  & 0.7 p.u.  & 20 ms \\
High voltage cease ($V_{\mathrm{HV}}$, $T_{\mathrm{HV}}$) & 1.15 p.u. & 500 ms \\
Double conv.\ ($\underline{V_{\mathrm{N}}}$, $T_{\mathrm{DC}}$) & 0.9 p.u.  & 500 ms \\
Low voltage reconnect ($V_{\mathrm{LVrecon}}$, $T_{\mathrm{recon}}$) & 0.9 p.u. & 500 ms \\
High voltage reconnect ($V_{\mathrm{HVrecon}}$, $T_{\mathrm{recon}}$) & 1.1 pu & 500 ms\\
Ramping time to rated power ($T_{\mathrm{ramp}}$) & - & 1 s\\
\hline
\end{tabular}
\end{table}

\begin{figure}[t]
  \centering
  \includegraphics[width=0.8\linewidth]{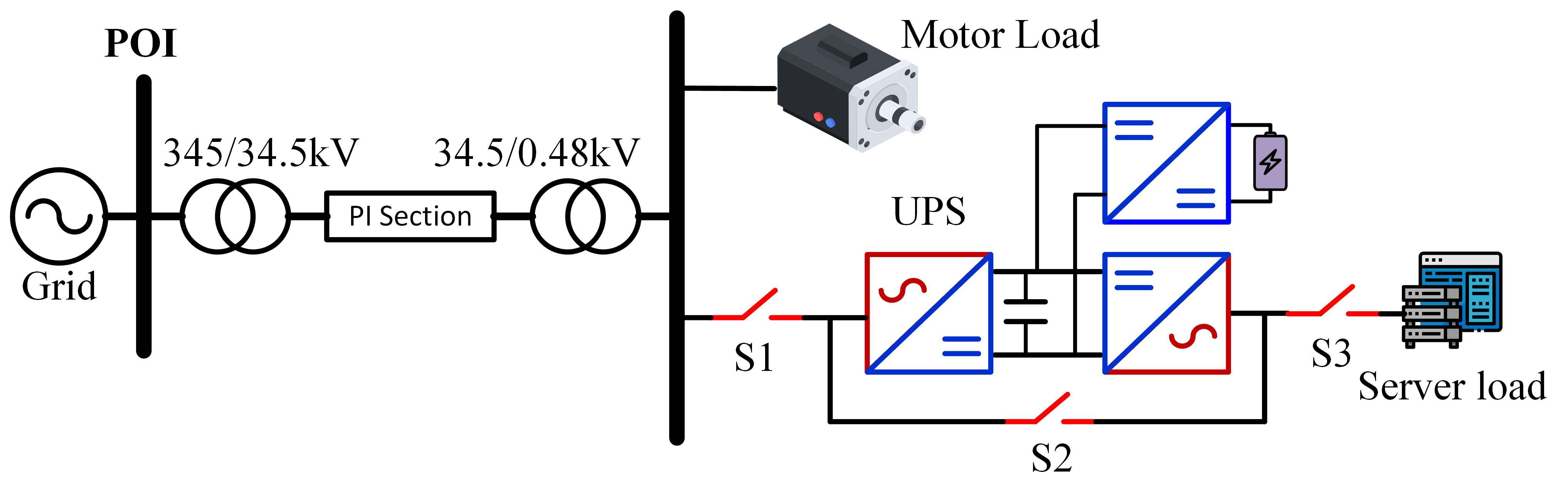}
  \caption{100~MW data center model.}
  \label{fig:dc100}
\end{figure}

\begin{figure}[t]
\centering
\includegraphics[width=0.9\linewidth]{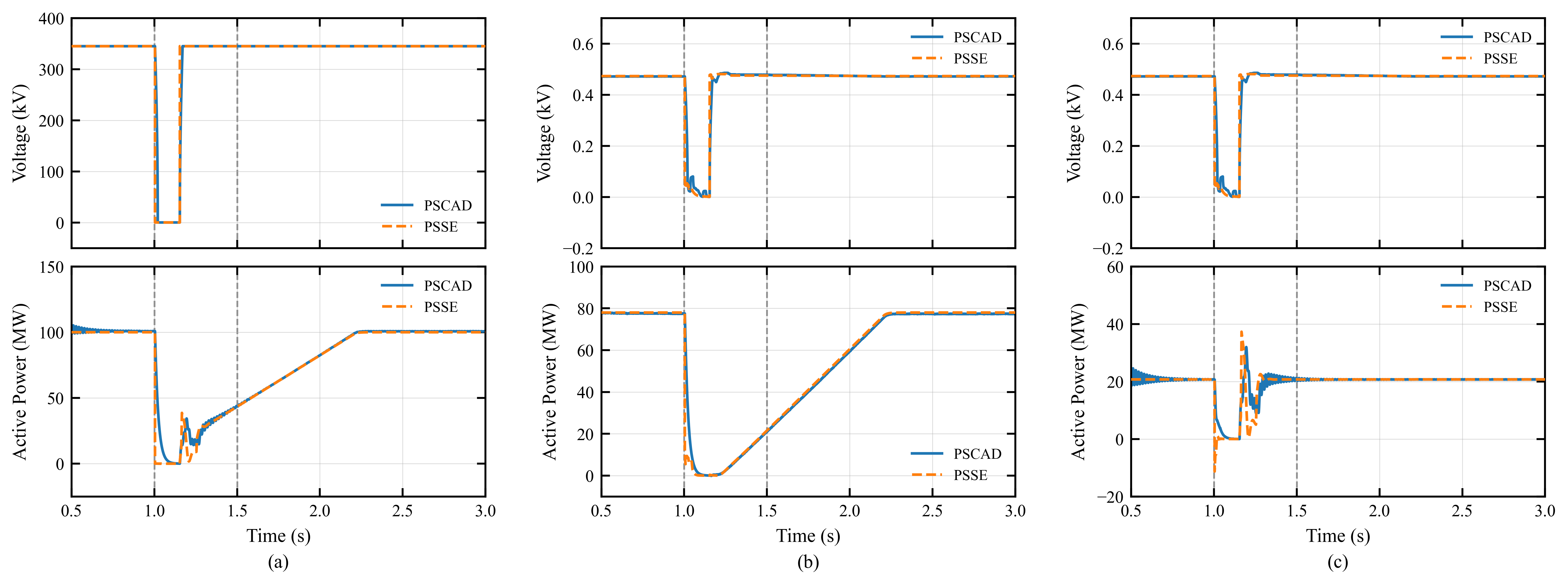}
\caption{Response of the 100~MW
data center to a three-phase grid fault:
(a) POI voltage and active power;
(b) IT load terminal voltage and active power consumption;
and (c) UPS terminal voltage and active power supplied to the IT load.}
\label{fig:voltage_dip}
\end{figure}

\begin{figure}[t]
  \centering
  \includegraphics[width=0.9\linewidth]{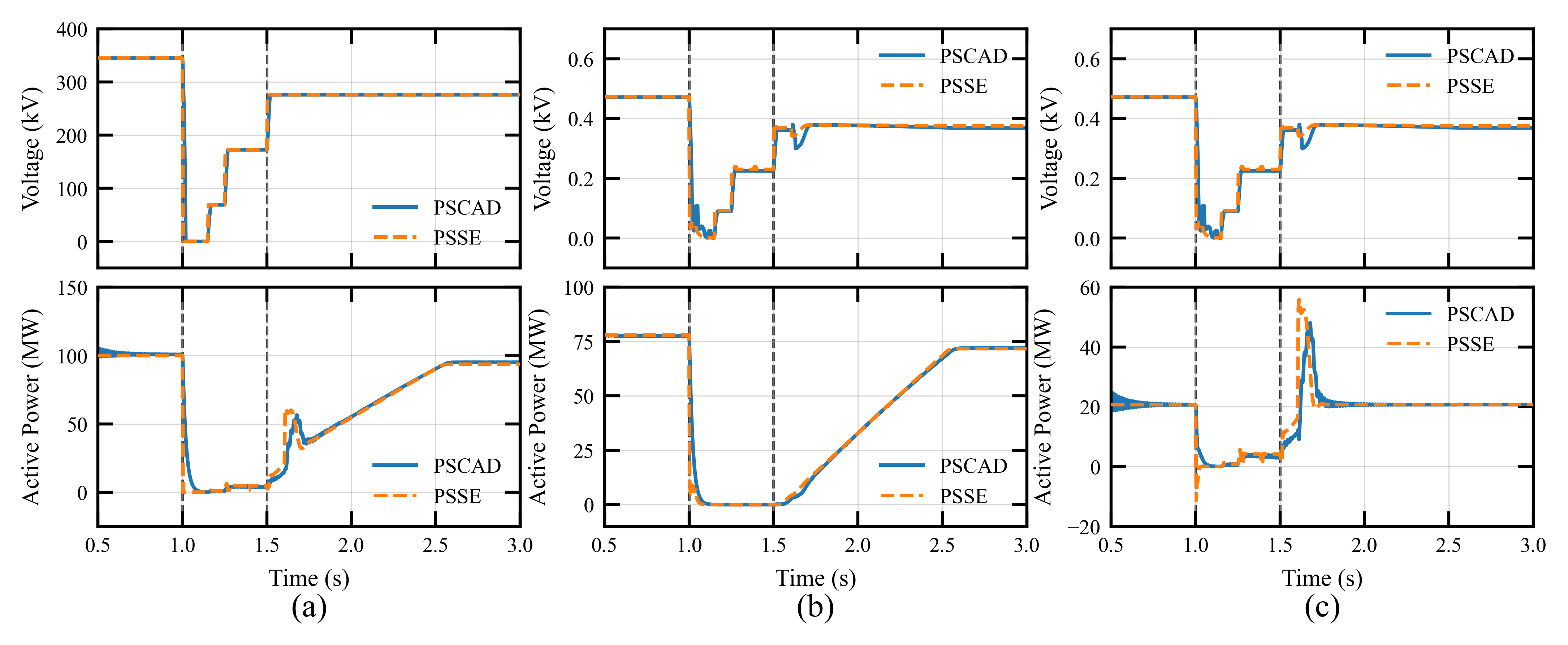}
  \caption{LVRT response of the 100~MW data center model: 
  (a) POI voltage and active power imported from the grid, 
  (b) server load terminal voltage and active power supplied by the grid, and
  (c) motor load terminal voltage and active power.}
  \label{fig:lvrt}
\end{figure}

\section{Simulation Results}
A scalable, parameterized EMT model of the data center is developed in PSCAD and benchmarked against the PSS\textregistered E PERC1 and CMPLDWG models for the IT and motor loads, respectively. Because PERC1 and CMPLDWG report only the voltage and active/reactive power at their aggregated terminals, the benchmark focuses on voltage and power at the UPS and motor-load terminals.

\subsection{Modeling and Validation of a 100~MW Data Center}
A 100~MW data center with 80\% IT load and 20\% motor load is modeled in PSCAD; its single-line diagram is shown in Fig.~\ref{fig:dc100}. Power is delivered from the grid through 345/34.5~kV and 34.5/0.48~kV step-down transformers. The model includes the UPS system, the server load, and an aggregated motor load. Lighting loads are omitted for brevity.

\subsubsection{Validation under a three-phase fault}
A symmetrical three-phase fault is applied at the POI at $t=1.0~\mathrm{s}$ and cleared at $t=1.15~\mathrm{s}$. Fig.~\ref{fig:voltage_dip} shows the resulting voltages and active power at the POI, IT load, and motor load. During the disturbance, the UPS draws no power from the grid, and the battery supplies the server load so that it keeps operating. This battery contribution is not shown in Fig.~\ref{fig:voltage_dip} because PERC1 does not report it and it therefore cannot be benchmarked. As the grid voltage recovers, the UPS ramps up its grid power at a set rate, while the battery supplies the rest of the server demand. The motors, in turn, are disconnected during the large voltage dip. When the voltage recovers, they reconnect and draw a burst of high power before settling back at their pre-fault level. Overall, the PSCAD/EMTDC and PSS\textregistered E results agree well at the POI, IT-load, and motor-load terminals.

\subsubsection{Validation under LVRT}
To validate the LVRT response of the 100~MW data center model, a symmetrical three-phase voltage sag is applied at $t=1.0~\mathrm{s}$ and cleared in stages, as shown in Fig.~\ref{fig:lvrt}. Before the disturbance, the POI voltage is about 345~kV, and the IT-load and motor-load terminal voltages are about 0.48~kV. During the fault, the voltages at all three locations drop close to zero and then recover in stages. The PSCAD/EMTDC and PSS\textregistered E voltage recovery profiles are nearly identical at the POI, IT-load terminal, and motor-load terminal.

At the IT load, the active power is initially about 78--80~MW and drops to nearly zero during the low-voltage interval. As the voltage recovers, the IT active power ramps up and settles at around 72--75~MW. The motor active power falls from about 20~MW to nearly zero during the fault and overshoots during reacceleration, peaking at about 50~MW in PSCAD/EMTDC and 55~MW in PSS\textregistered E before returning to about 20~MW. At the POI, the active power falls from about 100~MW to nearly zero and then recovers to about 95~MW. Overall, the two models agree well on the main LVRT voltage and power recovery behavior, while the EMT model also captures the fast transients associated with converter dynamics and motor reacceleration.

\begin{figure}[t]
\centering
\includegraphics[width=0.9\linewidth]{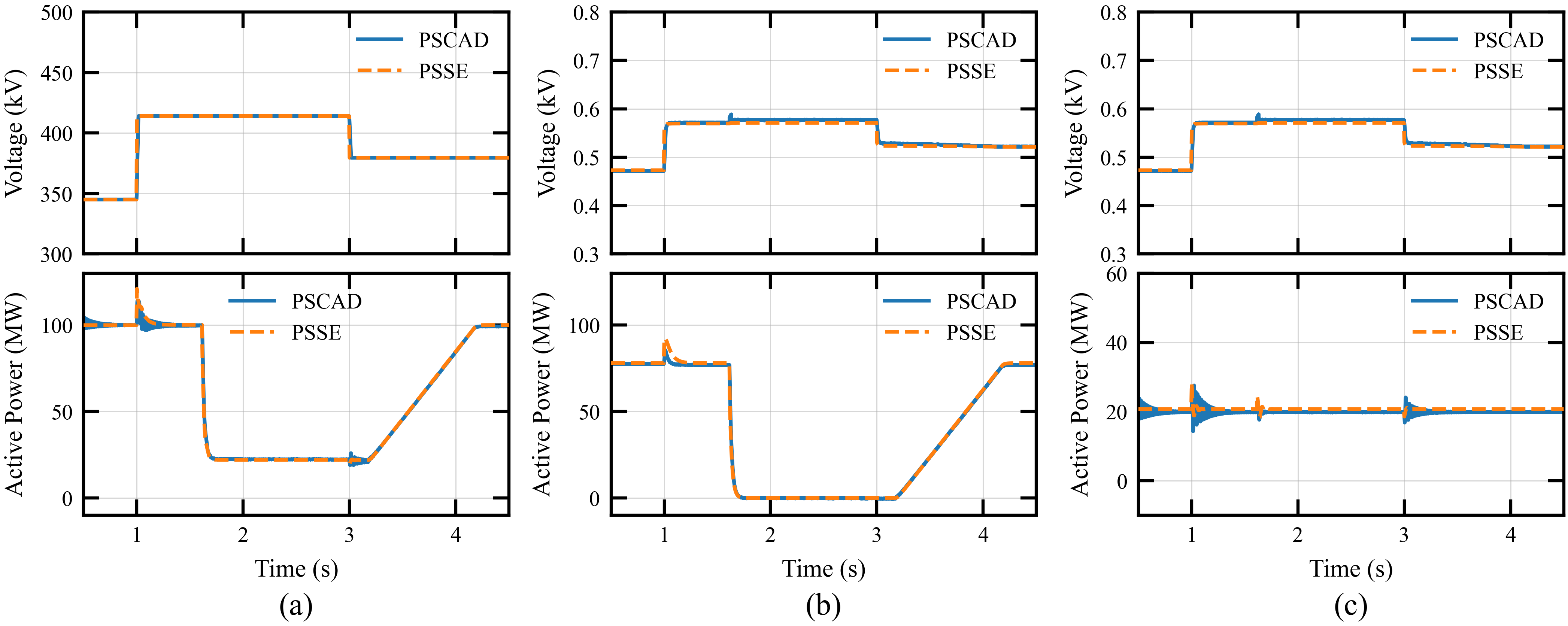}
\caption{HVRT response of the 100~MW data center model:
(a) POI voltage and active power imported from the grid,
(b) server load terminal voltage and active power supplied by the grid, and
(c) motor load terminal voltage and active power.}
\label{fig:hvrt}
\end{figure}

\subsubsection{Validation under HVRT}
To validate the HVRT response of the 100~MW data center model, a symmetrical three-phase voltage swell is applied at $t=1.0~\mathrm{s}$, as shown in Fig.~\ref{fig:hvrt}. Before the disturbance, the POI voltage is about 345~kV, and the IT-load and motor-load terminal voltages are about 0.48~kV. During the overvoltage event, the POI voltage rises to about 415~kV, and the low-voltage load terminal voltages rise to about 0.57--0.58~kV. After the event clears at around $t=3.0~\mathrm{s}$, the voltages return to their normal operating levels. The PSCAD/EMTDC and PSS\textregistered E voltage profiles are nearly identical at all three measurement locations.

At the IT load, the active power is initially about 78--80~MW and falls to nearly zero at around $t=1.6~\mathrm{s}$ during the ride-through interval. As the system recovers, the IT active power ramps back up and settles near 78~MW. The motor active power stays close to 20~MW throughout the event, with only short transients when the voltage changes. At the POI, the active power falls from about 100~MW to about 20~MW during the ride-through interval and then returns to about 100~MW. Overall, the PSCAD/EMTDC and PSS\textregistered E responses agree well under the HVRT event, confirming that the EMT model reproduces the benchmark voltage and active-power recovery behavior.

\section{Conclusion}
This paper presented an EMT dynamic model of a data center that captures the detailed behavior of its main subsystems, including the UPS, the cooling system, and the server load. A coordinated fault ride-through and mode-transition control strategy was implemented to keep the facility operating stably during grid disturbances. PSCAD/EMTDC simulations of a 100~MW facility, benchmarked against PSS\textregistered E, show that the model accurately captures both the steady-state and transient behavior of the data center. The results confirm that the facility operates stably under normal conditions, rides through faults reliably, and transitions smoothly between operating modes during grid disturbances.

\end{document}